\documentclass[
    twocolumn,
    superscriptaddress,
    amsmath, 
    amssymb,
    aps,
    reprint,
    floatfix
]{revtex4-2}
\usepackage{graphicx}

\usepackage[dvipsnames]{xcolor}
\usepackage{amsfonts}
\usepackage{bm}          
\usepackage{dcolumn}     
\usepackage{textcomp}    
\usepackage{multirow}
\usepackage{slashed}

\usepackage[caption=false]{subfig}

\usepackage{url}

\usepackage{hyperref}
\hypersetup{
  colorlinks=true,
  linkcolor=blue,
  citecolor=blue,
  filecolor=black,
  urlcolor=blue,
}

\definecolor{green2}{rgb}{0,0.56,0.32}

\def\figureautorefname~#1\null{Fig.\,#1\null}
\def\tableautorefname~#1\null{Tab.\,#1\null}
\def\equationautorefname~#1\null{Eq.\,(#1)\null}

\begin{document}

\preprint{APS/123-QED}

\title{\textbf{Probing compressed Higgsinos at the  FASER experiment } 
}%

\author{Shufang Su}
 \email{shufang@arizona.edu} 
\affiliation{Department of Physics, University of Arizona, Tucson, AZ 85721, USA}

\author{Wei Su}
 \email{suwei26@mail.sysu.edu.cn}
\affiliation{School of Science, Shenzhen Campus of Sun Yat-sen University,  Shenzhen 518107, P.R. China}
\affiliation{Institute of Theoretical Physics, Chinese Academy of
Sciences, Beijing 100190, P. R. China}

\author{Jin Min Yang}
\email{jmyang@itp.ac.cn}
\affiliation{Center for Theoretical Physics, Henan Normal University, Xinxiang 453007,  P. R. China}
\affiliation{Institute of Theoretical Physics, Chinese Academy of
Sciences, Beijing 100190, P. R. China}

\author{Pengxuan Zhu}
\email{pengxuan.zhu@adelaide.edu.au}
\affiliation{ARC Centre of Excellence for Dark Matter Particle Physics \& CSSM,\\ Department of Physics, University of Adelaide, Adelaide, SA 5005, Australia}

\author{Rui Zhu}
\email{zhurui@itp.ac.cn}
\affiliation{Institute of Theoretical Physics, Chinese Academy of
Sciences, Beijing 100190, P. R. China}

\date{\today}
\begin{abstract}
In the Minimal Supersymmetric Standard Model (MSSM), compressed Higgsinos spectrum ($\Delta m^0 \lesssim 1$ GeV) occurs when $|\mu| \ll |M_1|, |M_2|$ and ${\rm sign}(M_1\cdot M_2)<0$, which leads to a long-lived next-to-lightest neutralino. Such a long-lived neutralino could be copiously produced at the LHC, however escape the detection at the LHC main detectors. We examine the discovery potential at the FASER experiment and find that the FASER 2  could cover the neutral Higgsino mass up to about 130 GeV with mass splitting between 4 to 30 MeV. It is complementary to both the LHC Higgsino search in the $\Delta m^{0,\pm} \gtrsim 1$ GeV region, and displaced vertex and disappearing track searches of charginos with $\Delta m^\pm \lesssim 1$ GeV.
\end{abstract}

\maketitle


\noindent
\textbf{Introduction.}
Low-energy supersymmetry (SUSY) is one of the most compelling extensions of the Standard Model (SM) of particle physics, providing a natural solution to the hierarchy problem, leading to the unification of gauge couplings, and containing a natural dark matter (DM) candidate~\cite{Dreiner:2023yus, Martin:1997ns, Wang:2023suf, Haber:1984rc, Bae:2014yta}. However, the null result from the ongoing searches at the Large Hadron Collider (LHC) are pushing the SUSY parameter space to  the heavy masses or regions that escape the conventional collider searches.

The Higgsino mass parameter $\mu$ is unique in SUSY theory, as it is the only SUSY-conserving mass parameter in the Lagrangian.
In natural SUSY scenarios, the $\mu$ parameter should be close to the electroweak scale to avoid excessive fine-tuning in the electroweak symmetry breaking~\cite{Feng:2013pwa, vanBeekveld:2019tqp, Barbieri:1987fn, Arvanitaki:2013yja, Evans:2013jna, Baer:2014ica}. The Higgsino lightest supersymmetric particle (LSP) is a viable weakly interacting massive particle (WIMP) DM candidate~\cite{Delgado:2020url,Baer:2025srs}, with parameter space that remains consistent with current direct detection constraints~\cite{Martin:2024ytt, Martin:2024pxx}.   In the non-minimal SUSY models, light Higgsinos could play a unique role in the thermal history of the early Universe~\cite{Cao:2018rix, Cao:2019qng, Cao:2019evo, Cao:2021lmj, Cao:2021tuh, Yue:2025dqe, Liang:2025utb, Dong:2024juh, Wang:2024ozr, Chakraborti:2024pdn, Dai:2023pli, Khan:2025yit}. Other phenomenological studies on Higgsinos can be found in recent works~\cite{Araz:2025bww, Zhou:2025xol, Carpenter:2023agq, Baer:2025zqt, Baer:2024hqm, Nagata:2025ycf, Baer:2024tfo, Fowlie:2024nhs, Li:2023hey, VanBeekveld:2021tgn, Baer:2020sgm, Bae:2019dgg, Baer:2017pba, Wang:2016otm, Fan:2024jvz, Du:2023qvj, Wang:2023bmh, Li:2022zap, Wang:2022rfd, Wang:2018idz, Khan:2025azf, Yin:2024bwv, Fu:2023sfk, Agin:2024yfs, Rodd:2024qsi, Liu:2020muv, Liu:2020ctf, Zhao:2022pnv}. 

In the Higgsino-LSP scenario, searching for Higgsinos at the LHC faces several challenges. Due to the small mass differences 
$\Delta m^0 \equiv m_{\tilde{\chi}_2^0} - m_{\tilde{\chi}_1^0}$ and
$\Delta m^\pm \equiv m_{\tilde{\chi}_1^\pm} - m_{\tilde{\chi}_1^0}$ 
between Higgsinos, the decays of $\tilde{\chi}_2^0$ and $\tilde{\chi}_1^\pm$ produce very soft visible decay products. For $\Delta m^{0,\pm} \gtrsim 1$ GeV, typical Higgsino search channels include multi-soft leptons~\cite{ATLAS:2021moa, ATLAS:2019lng, CMS:2021edw, CMS-PAS-EXO-23-017}, monojet~\cite{ATLAS:2021kxv, Buanes:2022wgm, CMS:2021far, Agin:2023yoq}, or dijet channels~\cite{ATLAS:2024woy}.  
For the compressed Higgsino spectrum of $\Delta m^{0,\pm} \lesssim 1$ GeV, the next-to-lightest Higgsinos $\tilde{\chi}_1^{\pm}$ and $\chi_2^0$ could be long-lived particles(LLPs), which offering an alternative LLP mechanism in SUSY beyond commonly studied scenarios like $R$-parity violation~\cite{ATLAS:2015oan}, gauge-mediated SUSY breaking ~\cite{CMS:2024trg}, or split SUSY~\cite{CMS:2024trg}. Studies also show that the compressed neutralino is an important DM region~\cite{Martin:2007gf}. Recently, displaced vertex~\cite{ATLAS:2024umc, CMS:2025twk} and disappearing tracks~\cite{ATLAS:2022rme, CMS:2023mny} have been used to search for charged Higgsinos at the LHC with $\Delta m^\pm \lesssim 1$ GeV.  
Long-lived neutral Higgsinos $\tilde{\chi}_2^0$, however, remain unexplored in the current LHC experiments~\cite{ATLAS:2024woy}.
To study the compressed neutral Higgsinos with $\Delta m^0 \lesssim 1$ GeV, we propose a new detection channel, $\tilde{\chi}_2^0 \to \tilde{\chi}_1^0 \gamma$, using the LLP search at the ForwArd Search ExpeRiment (FASER). 

FASER is an active experiment at the LHC designed to detect LLP produced in the forward region of the ATLAS interaction point (IP). 
The LLPs travel in the very forward region,  and decay inside the FASER detector into  energetic particles~\cite{Feng:2017uoz,FASER:2018ceo, FASER:2018bac, FASER:2022hcn, FASER:2021ljd,FASER:2021cpr}. 
At the HL-LHC, the upgraded FASER 2 features a larger detection volume and improved sensitivity~\cite{FASER:2018eoc,Anchordoqui:2021ghd,Feng:2022inv}. 
Most current FASER and FASER 2 studies focus on light LLPs produced from the decays of $B$ and $K$ mesons, which are abundantly generated in the forward region of the LHC. To explore beyond the $B$ meson mass region, recent work primarily investigates decays from the $Z$ or Higgs boson~\cite{Helo:2018qej}. 
To fully explore the discovery potential of FASER 2, it is important to consider the production channels of electroweak-scale LLPs.

The production of Higgsinos at the LHC  via electroweak Drell–Yan process exhibits a characteristic enhancement along the beam direction. Furthermore, the longitudinal boost of the partonic collision system squeezes the polar angles and increases the $\tilde{\chi}_2^0$ energies along the beam axis, allowing them to decay inside the FASER 2 detection volume.
Therefore, the FASER 2 could be sensitive to a long-lived Higgsino with mass around electroweak scale. In the following sections, we analyze the FASER 2 discovery potential of the compressed Higgsino scenario in detail.

\vspace{0.5em}\noindent
\textbf{Electroweakino Sector.}
\label{sec:model}
In the MSSM, the supersymmetric partners of the electroweak gauge bosons  and the Higgsinos mix after the the electroweak symmetry breaking.  In the basis of $\psi_\alpha = (-i\tilde{B}, -i\tilde{W}^0, \tilde{H}_d^0, \tilde{H}_u^0)$, the symmetric neutralino mass matrix is written as  
\begin{equation}
\label{eq:massmatrix0}
\mathcal{M}_N =
	\begin{pmatrix}
	M_1 & 0 & -m_Z c_{\beta} s_{W} & m_Z s_{\beta} s_{W} \\
	& M_2 & m_Z c_{\beta} c_{W} & -m_Z s_{\beta} c_{W} \\
	 &  & 0 & -\mu \\
	 & & & 0
	\end{pmatrix}, 
\end{equation}
where $M_1$, $M_2$ and $\mu$ are the masses of Bino, Winos and Higgsinos, respectively, with $\theta_W$ being the Weinberg angle and $\tan\beta = v_u / v_d \equiv \langle {H}_u^0\rangle/\langle {H}_d^0\rangle$.  We use a shorthand notation of $c_x\equiv\cos x$ and $s_x\equiv\sin x$. The physical neutralino mass eigenstates $\tilde{\chi}_{i=1\ldots 4}^0$ are obtained by diagonalizing this mass matrix. 
The chargino sector consists of the charged Wino ($\tilde{W}^\pm$) and the charged Higgsinos ($\tilde{H}_u^+$, $\tilde{H}_d^-$), with chargino mass eigenstates $\tilde{\chi}_{1,2}^\pm$. At tree level, the electroweakino sector depends on four free  parameters: $M_1, M_2,  \mu$, and $\tan{\beta}$.

In the Higgsino-LSP scenario ($|\mu| \ll |M_1|, |M_2|$), $\tilde{\chi}_{1,2}^0$ and $\tilde{\chi}_1^\pm$ are Higgsino-like with small mass differences of a few GeV. Treating the $m_Z$-dependent terms in Eq.~(\ref{eq:massmatrix0}) as perturbations, the masses of the two light neutralino states are approximated given by (assuming \(\mu > 0\) in the following discussion):
\begin{equation}
\label{eq:mnue}
m_{\tilde{H}_{1,2}^0} \approx \mu \pm \frac{m_Z^2}{2}\left(1 \mp s_{2\beta} \right) \left(
\frac{c^2_W} { M_2 \pm \mu } + \frac{ s^2_W}{ M_1 \pm \mu } \right). 
\end{equation}
The relative signs of the three mass parameters, $\mu$, $M_1$, and $M_2$,  affect the mass corrections of the Higgsino-like states. Therefore, either of $\tilde{H}_{1,2}^0$ can be the LSP. The mass splitting between the two neutralinos is given by  
\begin{equation}
\label{eq:msplitn2n1}
\Delta m^0 \approx
m_Z^2 \left| \frac{c_W^2(M_2 + \mu s_{2\beta})}{M_2^2 - \mu^2} + \frac{s_W^2( M_1 + \mu s_{2\beta})}{M_1^2 - \mu^2 } \right|.
\end{equation}
For $|M_{1,2}|\gg m_Z$,  the mass splitting $\Delta m^0 $ is typically about 1 GeV or less. The dominant decay channels are 
\begin{equation}
\tilde{\chi}_2^0 \to \tilde{\chi}_1^0 + \gamma,\ \ \tilde{\chi}_2^0 \to \tilde{\chi}_1^0 Z^* \to \tilde{\chi}_1^0 f\bar{f}.
\end{equation}
Radiative decay has been investigated in Refs.~\cite{Haber:1988px, Gunion:1987yh, Ambrosanio:1995az, Ambrosanio:1996gz, Baer:2002kv}, and implemented in \texttt{SDECAY}~\cite{Djouadi:2006bz}:
\begin{equation}
\label{eq:Gchi20}
	\Gamma(\tilde{\chi}_2^0 \to \tilde{\chi}_1^0 \gamma) = \frac{g_{\tilde{\chi}_2^0 \tilde{\chi}_1^0 \gamma}^2 }{8 \pi } \frac{\left( m_{\tilde{\chi}_2^0}^2 - m_{\tilde{\chi}_1^0}^2 \right)^3}{m_{\tilde{\chi}_2^0}^5},
\end{equation}
with $g_{\tilde{\chi}_2^0 \tilde{\chi}_1^0 \gamma} \propto e g^2 /16\pi^2 $  being an effective coupling generated via one-loop diagrams. 
For $\Delta m^0\sim 1$ GeV, ${\rm Br}(\tilde{\chi}_2^0 \rightarrow \tilde{\chi}_1^0 Z^*)$ is less than  30\% for a sub-TeV Higgsino, and reduces to less than $1\%$ for $\Delta m^0 \lesssim  0.1~{\rm GeV}$.

 Similarly, the mass splitting between the Higgsino-like chargino and the LSP $\tilde{\chi}_1^0$ is
 {\small
\begin{equation}
\label{eq:msplitc1N1}
\Delta m^\pm  = \mp {\rm sign}(\mu) \frac{m_Z^2}{2} \left( \frac{ c_W^2 (1 \pm s_{2\beta})}{M_2 \mp \mu }  +  \frac{ s_W^2 (1\mp s_{2\beta})}{M_1 \pm \mu} \right). 
\end{equation}
}
The two values correspond to the cases where $\tilde{H}_1^0$ or $\tilde{H}_2^0$ is the lighter neutralino, respectively. When $\Delta m^\pm$ is a few hundred MeV, the charginos $\tilde\chi_1^\pm$ can be long-lived. For $\Delta m^\pm \gtrsim m_\pi$, $\tilde\chi_1^\pm \to \tilde\chi_1^0 \pi^\pm$ is strongly phase-space suppressed, leading to a proper decay length $c\tau$ of order centimeters to tens of centimeters. For $\Delta m^\pm < m_\pi$, charginos decay via $\tilde\chi_1^0 \ell^\pm \nu$, resulting in much longer $c\tau$ values, ranging from tens of centimeters to meters. Such long-lived charginos give rise to characteristic collider signatures of displaced vertices and disappearing tracks~\cite{Chen:1996ap, ATLAS:2024umc, CMS:2025twk, ATLAS:2022rme, CMS:2023mny}.

\begin{figure}[t]
	\centering
    \makebox[\linewidth][c]{
    \includegraphics[width=0.495\linewidth]{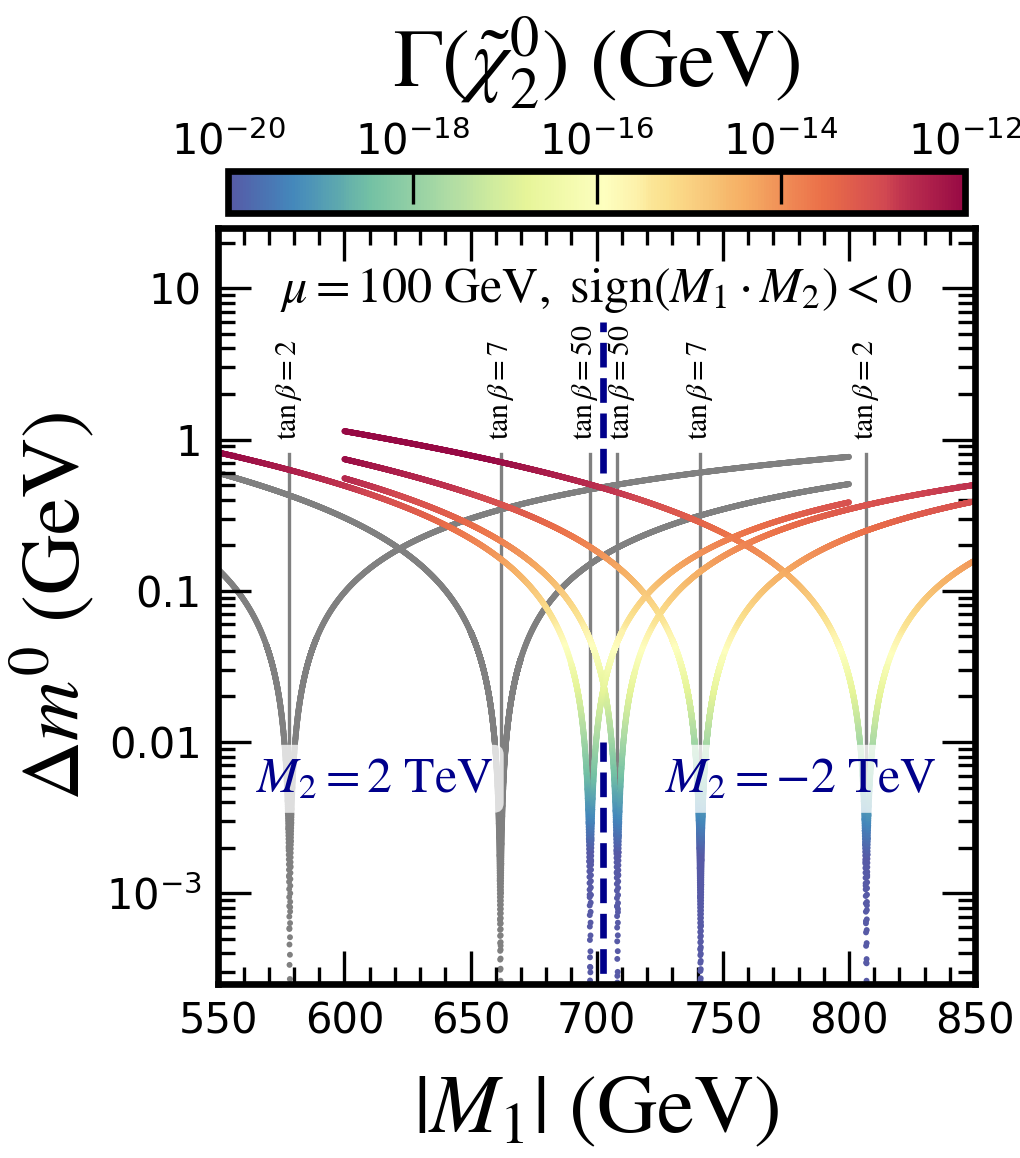}
	\includegraphics[width=0.495\linewidth]{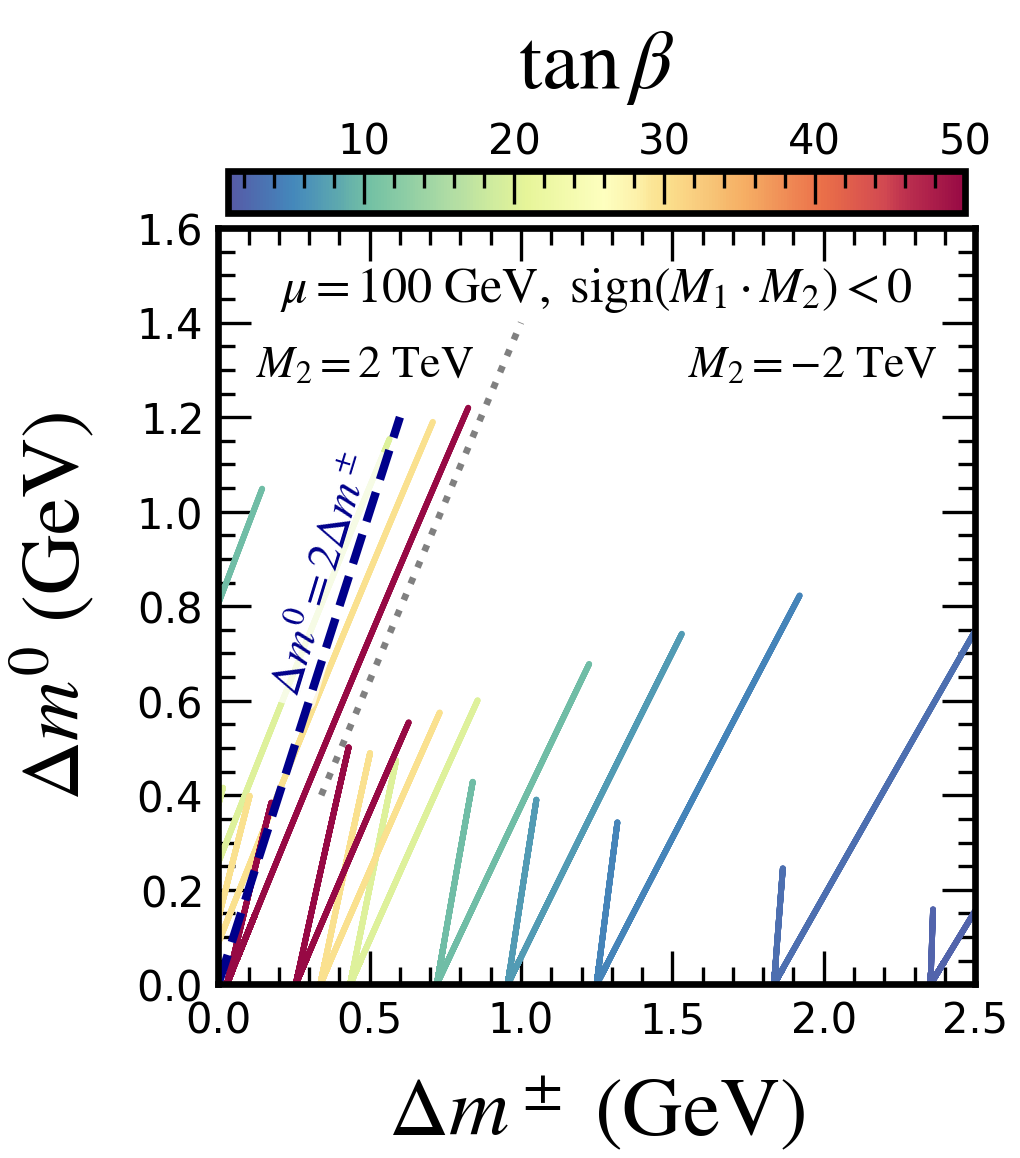}
    }
    \vspace{-.8cm}
	\caption{\label{fig:para} The samples in the $(\left| M_{1}\right|, \Delta m^{0})$ plane (left) and $(\Delta m^{\pm}, \Delta m^{0})$ plane (right), color-coded by $\Gamma_{\tilde{\chi}^0_2}$ (left) and $\tan\beta$ (right), respectively, for $M_2=\pm 2$ TeV and $\mu=100$ GeV. Gray points denote charged Higgsino LSPs.   
    }
\end{figure}

A highly compressed Higgsino spectrum occurs when 
\begin{equation}\label{eq:sim}
	\frac{M_1}{M_2}  \approx - \tan^2{\theta_W}. 
\end{equation}
The leading term inside the brackets of  Eq.~(\ref{eq:msplitn2n1}) and Eq.~(\ref{eq:msplitc1N1}) cancels out.  $\Delta m^0$ could reach 1 GeV or smaller, resulting in a long-lived Higgsino $\tilde\chi_2^0$.
In the left panel of Fig.~\ref{fig:para}, we show the $\Delta m^0$ dependence on $|M_1|$ for $M_2 = \pm 2$ TeV, $\mu=100$ GeV and ${\rm sign}(M_1\cdot M_2)<0$, 
using the \texttt{SUSYHIT} package. The dip structure in $M_1$ corresponds to cancellation between the  $M_1$ and $M_2$ terms in Eq.~(\ref{eq:msplitn2n1}). The color map shows the partial decay width $\Gamma(\tilde{\chi}_2^0)$.  Near these cancellation points, $\Gamma(\tilde{\chi}_2^0)$ could be $10^{-16}$ GeV or smaller, which corresponding to $c\tau$ of 2 meter or larger. 
The location of the dip depends on the value of $\tan\beta$, as shown in Eq.~(\ref{eq:msplitn2n1}). Note that for ${\rm sign}(M_2\mu)>0$,  $\Delta m^\pm$ in Eq.~(\ref{eq:msplitc1N1}) is mostly negative except for large $\tan\beta$, which corresponds to a chargino LSP scenario that is disfavored by dark matter considerations, as indicated in gray in the figure.

Loop corrections to electroweakino mass matrices primarily shift the mass matrix entries. Including one-loop mass corrections would shift the exact cancellation ($M_1$, $M_2$) relation but our general conclusion of an extremely compressed Higgsino spectrum still holds.

The right panel of Fig.~\ref{fig:para} shows $\Delta m^\pm$ versus $\Delta m^0$ panel with $\tan{\beta}$ indicated by the heat map. Each broken line contains two line segments with different slopes, for $\tilde{H}_{1}^0$ or $\tilde{H}_2^0$ being the LSP, as described in Eq.~(\ref{eq:mnue}).
The dotted line separates the region of $M_2=\pm 2$ TeV. In the general MSSM parameter space of Higgsino LSP with $\Delta m^0 > 1~{\rm GeV}$, 
$\Delta m^0 \approx 2 \Delta m^\pm$ holds approximately, as demonstrated by  the leading terms of Eqs.~(\ref{eq:msplitn2n1}) and (\ref{eq:msplitc1N1}). However, in the compressed regime ($\Delta m^0 \lesssim 1~{\rm GeV}$) studied in this paper, this relation no longer holds. $\Delta m^\pm$ could be in the range of a few GeV while $\Delta m^0$ remains small. This has important implication on the current experimental search limits, as explained in the next section.

\begin{figure}[t]
	\centering
	\includegraphics[width=\linewidth]{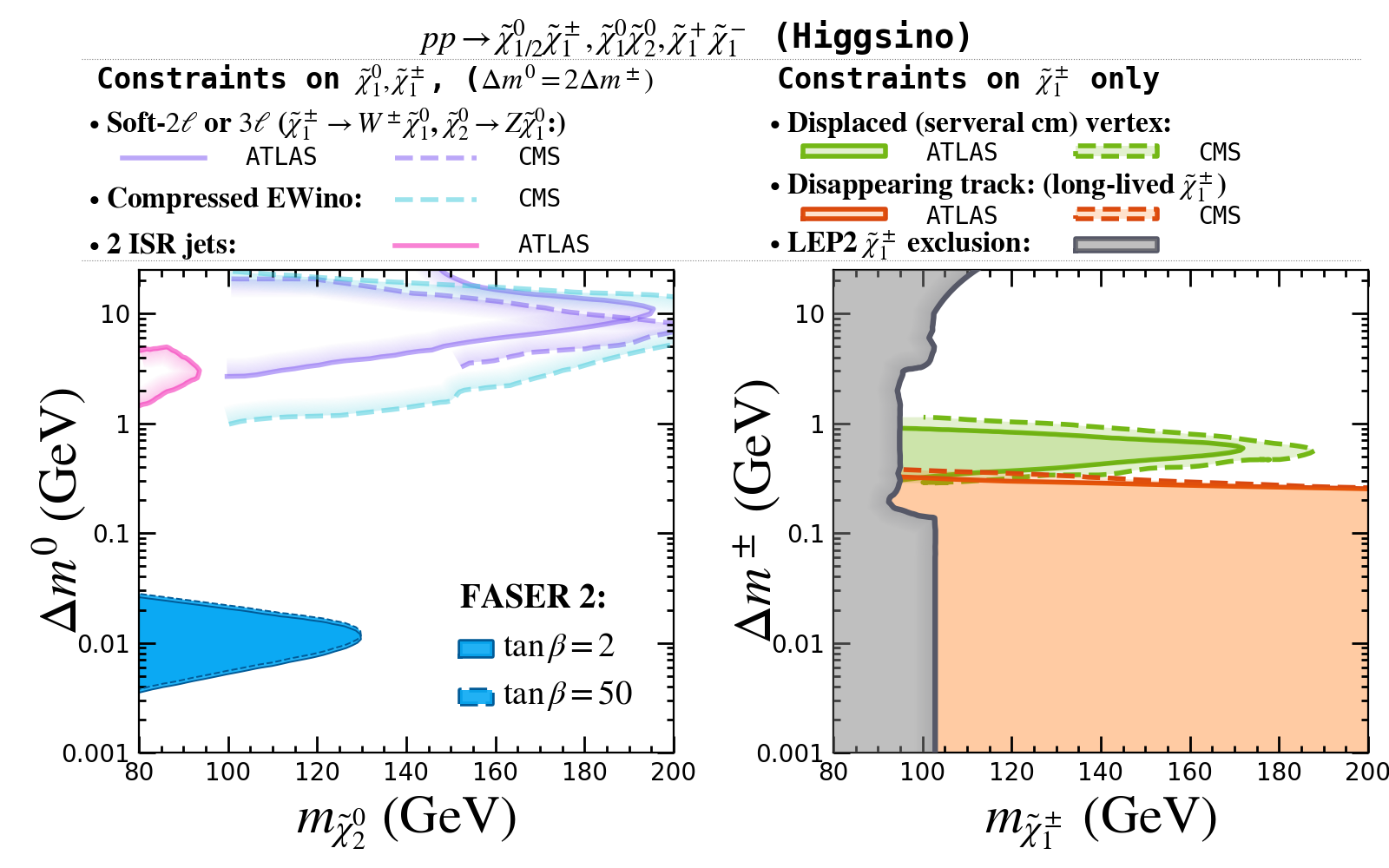}
    \vspace{-.8cm}
\caption{\label{fig:expconstrain} 
Experimental constraints on the Higgsino-like electroweakinos in the plane of mass splittings versus the mass of the neutralino (left) and chargino (right).  Projected sensitivities from FASER~2 are shown for $\tan\beta = 2$ (solid) and $\tan\beta = 50$ (dashed) in the left panel. 
}
\end{figure}

\vspace{0.5em}\noindent
\textbf{Experimental Constraints.}
\label{sec:exp}
%
Extensive searches for Higgsino-like electroweakinos have been conducted at both LEP and the LHC. Figure~\ref{fig:expconstrain} shows the 95\% C.L. exclusion limits in the neutralino (left panel) and chargino (right panel) mass vs. mass splitting planes.
$\tilde{\chi}^+_1\tilde{\chi}^-_1$ pair production has been studied at the LEP across a broad range of mass splittings $\Delta m^\pm$, using ISR-photon tagging with soft decay products, and displaced or quasi-stable tracks for the compressed spectrum (gray region in the right panel)~\cite{Abdallah:2002aik,DELPHI:1999jsl,ALEPH:2002gap,L3:1999twz,DELPHI:1997tcq}.
Both ATLAS and CMS have searched for nearly degenerate Higgsinos in various channels, including multilepton final states with missing transverse momentum (purple)~\cite{ATLAS:2021moa, ATLAS:2019lng, CMS:2021edw}, 
compressed electroweakino searches (cyan)~\cite{CMS-PAS-EXO-23-017}, and ISR-assisted search (pink)~\cite{ATLAS:2024woy}. 
Searches specific to charginos include low-momentum mildly displaced vertex track searches (green)~\cite{ATLAS:2024umc, CMS:2025twk} and disappearing track searches (orange)~\cite{ATLAS:2022rme, CMS:2023mny}). The left panel shows that current LHC searches for nearly degenerate $\tilde{\chi}^0_2$ states lose sensitivity when $\Delta m^0 \lesssim 1~\rm{GeV}$.  The right panel shows that regions of $\Delta m^\pm \lesssim 1~\rm{GeV}$ are tightly constrained by searches for relatively long-lived charginos.

Note that while the multilepton channel (purple lines) and compressed electroweakino search (cyan lines) exclude a large region of Higgsino parameter space for $\Delta m^0 \gtrsim 1$ GeV, those limits are under the assumption of the mass relation $\Delta m^0 = 2 \Delta m^\pm$. In the compressed Higgsino mass region, this mass relation no longer holds, as indicated in the right panel of Fig.~\ref{fig:para}. Therefore, those search limits would be relaxed in the compressed Higgsino scenario under current study. In particular,  $\Delta m^\pm \gtrsim 1$ GeV are unconstrained since the corresponding $\Delta m^0$ could be very small.



\vspace{0.5em}\noindent
\textbf{FASER Sensitivity to LLP $\tilde{\chi}_2^0$.}
\label{sec:faser}
%
The FASER experiment~\cite{Feng:2022inv, FASER:2018eoc} is designed to search for LLPs produced in the far-forward region of the LHC. 
The detector is a compact cylindrical decay volume with radius $R = 0.1~\mathrm{m}$ and length $D = 1.5~\mathrm{m}$, located 480 $\mathrm{m}$ downstream of the ATLAS IP. 
FASER~2 is the upgrade of FASER at the HL-LHC, which is located $L=620~\mathrm{m}$ downstream of the ATLAS IP with a cylindrical decay volume of $R=1~\mathrm{m}$ and $D=10~\mathrm{m}$.
Moderate variations of the detector dimensions within the proposed design range~\cite{Feng:2022inv, FASER:2018eoc} do not qualitatively affect the sensitivity of FASER 2 to compressed Higgsino scenario. 

For a LLP of mass $m$ and decay lifetime $\tau$, produced at the IP with momentum $p$ and polar angle $\theta$, the probability of decaying inside the fiducial volume of FASER 2  is approximately
\begin{equation}\label{eq:FASERprob}
    \mathrm{Prob}_{\rm FASER} \simeq \frac{D}{\lambda} e^{-L/\lambda} ~ \Theta \left(R - L\tan\theta\right),
\end{equation}
where  
$ \lambda = c\tau\,\beta\gamma $
is the mean LLP decay length in the lab frame, and $\Theta(x)$ the Heaviside step function enforcing the condition that the particle’s trajectory enters the detector before decaying.

\begin{figure}[tbp]
     \centering
     \includegraphics[width=0.75\linewidth]{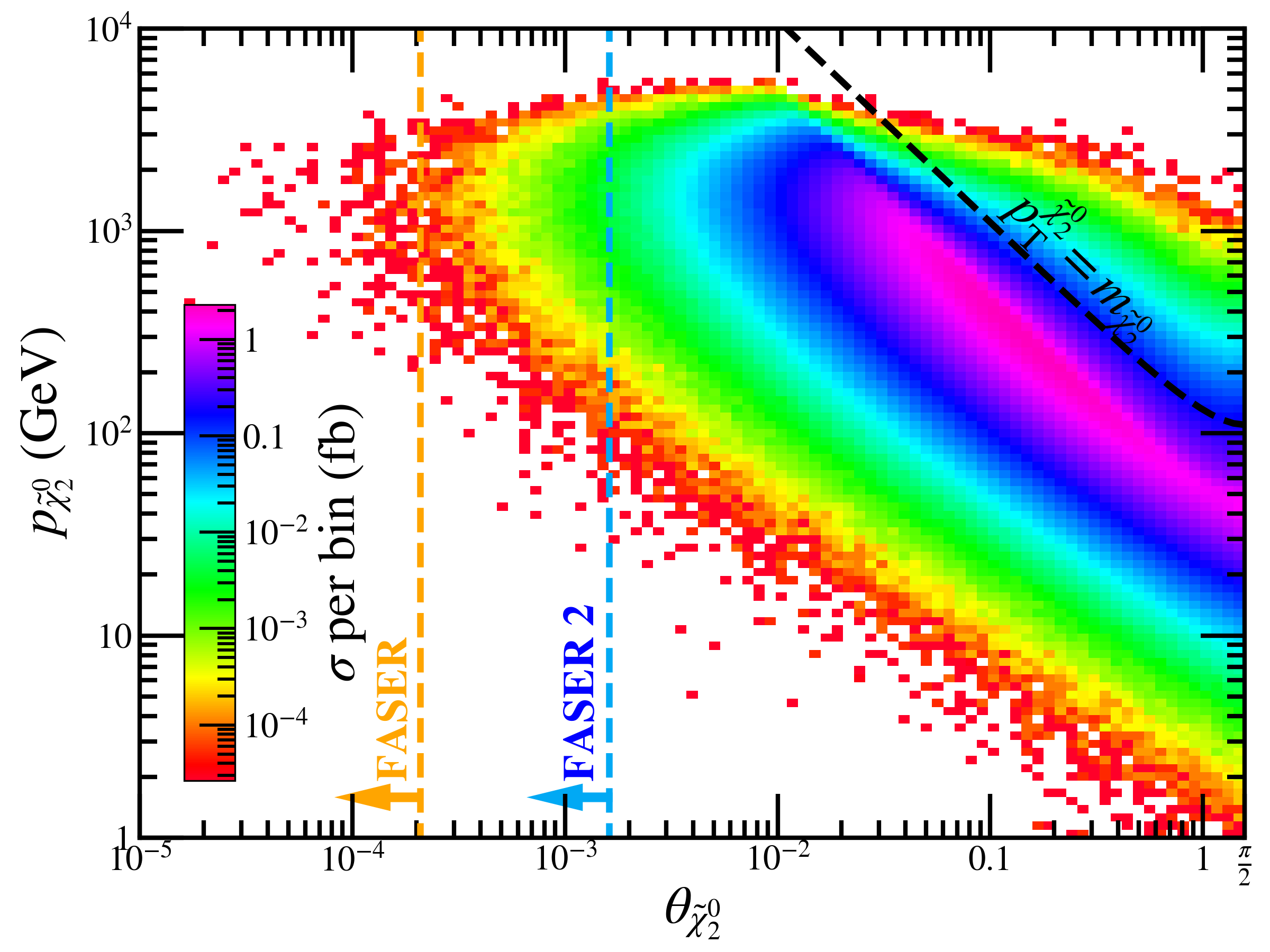}
     \vspace{-.3cm}
 \caption{\label{fig:xsect} 
 Density map of $p_{\tilde{\chi}_2^0}$ versus $\theta_{\tilde{\chi}_2^0}$ for $\tilde{\chi}_2^0$ production at the LHC, shown for $\mu = 110~\mathrm{GeV}$. 
The color scale represents the differential cross section per bin in unit of fb. 
Vertical dashed lines indicate the angular acceptances of FASER (orange) and FASER~2 (blue), and the black dashed curve shows $p^{\tilde{\chi}_2^0}_{\rm T} = m_{\tilde{\chi}_2^0}$.
 }
\end{figure}

Typical benchmark scenarios considered in the FASER study~\cite{FASER:2018eoc} involve light LLPs originating from meson decays in the forward region. 
However, the compressed Higgsino scenario examined here is characterized by an electroweak Drell-Yan production~\cite{Haber:1984rc}.  
For the numerical analysis, we generate all Higgsino pair–production channels using \texttt{Pythia8.3}~\cite{Bierlich:2022pfr}, restricting to phase space of hard scattering transverse momentum less than 50 GeV~\cite{Skands:2014pea}.   Fig.~\ref{fig:xsect} shows the $(\theta,p)$ distribution for $\tilde{\chi}_2^0$ production ($\mu = 110~\mathrm{GeV}$).  
$\tilde{\chi}_2^0$ is highly boosted and strongly forward-peaked, with the differential cross section enhanced at small polar angles. 
The larger transverse size of FASER~2 allows substantial overlap with the forward Higgsino flux. 
The momenta typically range from a few hundred GeV to multi-TeV.

The expected 3 event FASER 2 detection reach of a long-lived neutralino $\tilde{\chi}_2^0$ was shown in the left panel of Fig.~\ref{fig:expconstrain} as blue regions, for an integrated luminosity of 3 ${\rm ab}^{-1}$.  For $80<m_{\tilde\chi_2^0}<130$ GeV, FASER 2 is sensitive to  $\Delta m^0$ ranges from 4 to 30 MeV. 
The FASER 2 reach has little dependence on the value of $\tan\beta$, as illustrated by the blue solid (dashed) curve for $\tan\beta=2$ (50).

\begin{figure}[tbp]
\centering
    \makebox[\linewidth][c]{
	\includegraphics[width=0.5\linewidth]{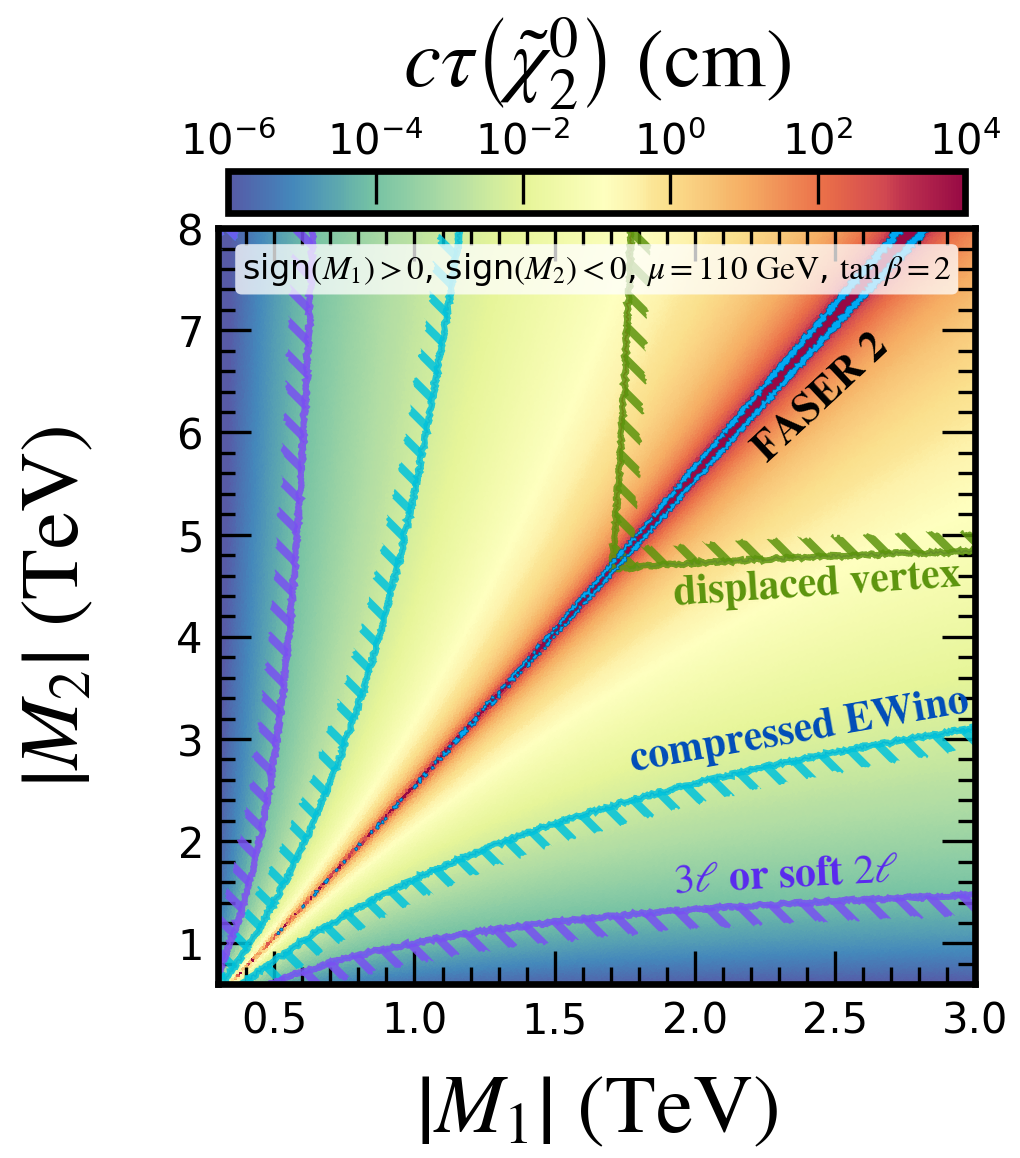}	
    \includegraphics[width=0.5\linewidth]{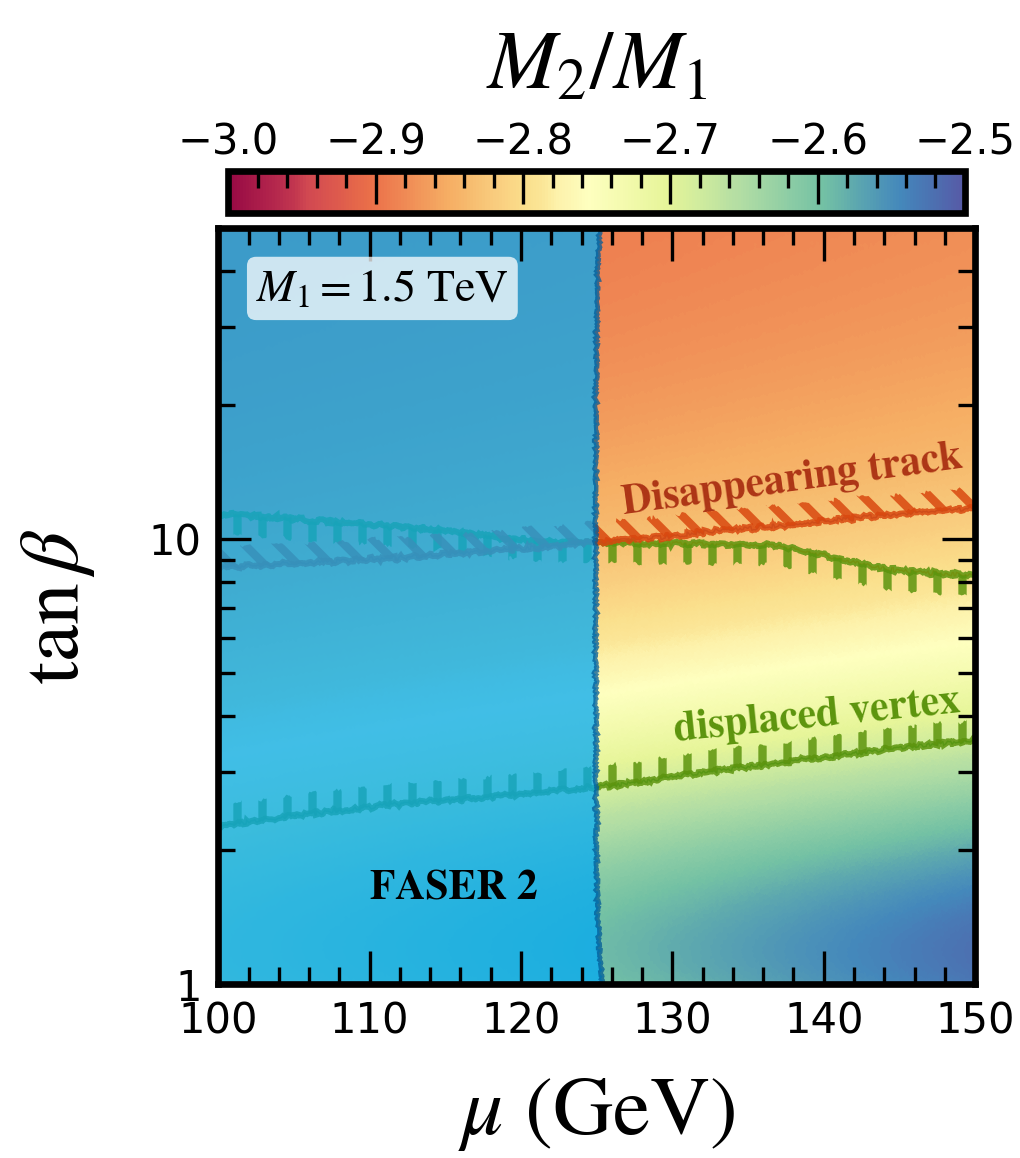}
    }
    \vspace{-.8cm}
\caption{\label{fig:ps} \textit{Left}: Coverage of LHC searches and FASER 2 in the $(|M_1|,|M_2|)$ plane, with heat map indicating the proper lifetime of $\tilde{\chi}_2^0$. 
\textit{Right}: Projected sensitivity of the LHC and FASER 2 in the $(\mu,\tan\beta)$ plane, with the color indicating the value of the $M_2/M_1$ ratio that yields the maximal sensitivity of FASER 2.
}
\end{figure}

Given the dependence of  $\Delta m^0$ and $\Delta m^\pm$ on the MSSM gaugino mass parameters $M_1$ and $M_2$, the left panel of Fig.~\ref{fig:ps} shows the FASER 2 sensitivity (blue region) in the $(|M_1|,|M_2|)$ plane for a benchmark $\mu = 110~\mathrm{GeV}$ and $\tan\beta=2$. Also shown are the LHC search limits from multilepton searches~\cite{ATLAS:2021moa, ATLAS:2019lng, CMS:2021edw}, compressed electroweakino searches~\cite{CMS-PAS-EXO-23-017}, and displaced-track searches for intermediate-lifetime charginos~\cite{ATLAS:2024umc, CMS:2025twk}.

In the region of large $|M_1|$ and $|M_2|$, $\Delta m^\pm$ becomes small, leading to a relatively long-lived chargino  excluded by the disappearing-track or displaced vertex searches.
When either $M_1$ or $M_2$ is relatively small, $\Delta m^0$ and $\Delta m^\pm$ is around a few GeV. Such regions are excluded by the multilepton  and compressed electroweakino searches. 
In regions of parameter space where the chargino is not long-lived while the neutralino mass splitting $\Delta m^0$ remains small, conventional LHC searches lose sensitivity. However,  $\tilde\chi_2^0$  can be long-lived and highly boosted, making FASER~2 uniquely sensitive to this otherwise unconstrained region. For $\mu = 110~\mathrm{GeV}$, FASER~2 probes  $\Gamma_{\tilde{\chi}_2^0}$ in the range of $(4.7\times10^{-19} - 1.2\times10^{-17})~\mathrm{GeV}$, corresponding to two blue thin regions close to the diagonal direction.

The right panel of Fig.~\ref{fig:ps} shows the projected sensitivity in the $(\mu,\tan\beta)$ plane for $M_1=1.5~\mathrm{TeV}$, with the heat map indicating the $M_2/M_1$ ratio that yields the maximal FASER 2 sensitivity.
FASER~2 covers $\mu < 125~{\rm GeV}$ with only a slight dependence on $\tan\beta$.
However, the LHC's sensitivity to chargino searches depends strongly on $\tan\beta$. The displaced-vertex searches for intermediate lifetime charginos~\cite{ATLAS:2024umc, CMS:2025twk} exclude $2\lesssim \tan\beta \lesssim 10$, while the disappearing-track searches targeting long-lived charginos~\cite{ATLAS:2022rme, CMS:2023mny} cover $\tan\beta \gtrsim 9$. 

Our analyses reveals the complementarity between LHC searches and FASER~2 reach in the MSSM electroweakino parameter space.
Conventional LHC searches probe regions with sizable mass splittings $\Delta m^{0,\pm}\gtrsim 1$ GeV or relatively long-lived charginos with $\Delta m^\pm \lesssim 1$ GeV, while FASER~2 is sensitive to small $\Delta m^0$ around 4 - 30 MeV.
Note that the FASER~2 sensitive regions correspond to a restricted range of the gaugino mass ratios $M_2/M_1$. A positive signal from FASER 2 not only points to a natural SUSY spectrum but also provides valuable insight into the structure of SUSY breaking at high scales.

\vspace{0.5em}\noindent
\textbf{Conclusions.}
\label{sec:sum}
Within the SUSY framework, the compressed electroweakino spectrum  with $\Delta m^{0,\pm}\lesssim 1$ GeV is challenging to be explored experimentally. While the displaced vertices and disappearing tracks provide useful probes for the chargino  sector with $\Delta m^\pm \lesssim 1 $ GeV, regions with small $\Delta m^0$ remain undetected by the LHC main detectors.
In our study, we focused on a specific region of the Higgsino LSP scenario ($|\mu| \ll |M_{1,2}|$) with $M_2/M_1 \approx -\tan^2{\theta_W}$, where the Higgsino-like neutralinos $\tilde{\chi}_1^0$ and $\tilde{\chi}_2^0$ become nearly degenerate with $\Delta m^0 \lesssim 1$ GeV, resulting in a long-lived $\tilde{\chi}_2^0$. Although this scenario is challenging at the conventional LHC searches, the upcoming FASER 2 experiment offers unique detection sensitivities. FASER 2 could probe $|\mu|$ up to about $125~{\rm GeV}$ and a mass splitting $\Delta m^0$ between 4 to 30 MeV, corresponding to a decay width $\Gamma_{\tilde{\chi}_2^0}$ of approximately $10^{-19} - 10^{-17}~{\rm GeV}$. FASER 2 and conventional LHC searches are highly complementary, together covering a significant portion of the Higgsino LSP scenario parameter space.




While FASER 2 is designed to detect the LLPs at the forward region, it is typically sensitive to light particles copiously produced via the decay of mesons.
Our study demonstrated the sensitivity of FASER 2 to the LLP with masses at the electroweak scale, which opens the door to explore other electroweak scale long-lived new particles, beyond what is currently studied from $B$ meson decays, or via the rare invisible decay of the SM-like Higgs, or at future lepton collider~\cite{Blondel:2022qqo}. 


\vspace{0.5em}\noindent
\textbf{Acknowledgments.}
SS is supported by the Department of Energy under Grant No. DE-
FG02-13ER41976/DE-SC0009913. 
WS is supported  by the Natural Science Foundation of China (NSFC) under Grant No. 12305115. 
JMY is supported by NSFC under Grant No. 12335005 and by PI Research Fund under Grant No. 5101029470335. 
PZ is supported by the ARC Discovery Project DP220100007 and by the Center for the Subatomic Structure of Matter (CSSM).


\bibliographystyle{CitationStyle}
\bibliography{ref_type1}
\end{document}